\documentclass[sigconf]{acmart}
\usepackage{amsmath}
\usepackage{graphicx} 
\usepackage{subcaption} 
\usepackage[dvipsnames]{xcolor}
\AtBeginDocument{%
  }

\newcommand{\novelty}[1]{{\color{black}{#1}}}
\newcommand{\expert}[1]{{\color{black}{#1}}}
\newcommand{\methods}[1]{{\color{black}{#1}}}
\newcommand{\clarity}[1]{{\color{black}{#1}}}

\copyrightyear{2026}
\acmYear{2026}
\setcopyright{cc}
\setcctype{by}
\acmConference[CHI '26]{Proceedings of the 2026 CHI Conference on Human Factors in Computing Systems}{April 13--17, 2026}{Barcelona, Spain}
\acmBooktitle{Proceedings of the 2026 CHI Conference on Human Factors in Computing Systems (CHI '26), April 13--17, 2026, Barcelona, Spain}
\acmPrice{}
\acmDOI{10.1145/3772318.3791498}
\acmISBN{979-8-4007-2278-3/2026/04}

\begin{document}

\title[Expert Perspectives on Teen-Centered Social Media Risk Detection Technologies]{From ``Fail Fast'' to ``Mature Safely:'' Expert Perspectives as Secondary Stakeholders on Teen-Centered Social Media Risk Detection}

\author{Renkai Ma}
\affiliation{%
  \institution{University of Cincinnati}
  \city{Cincinnati}
  \state{Ohio}
  \country{USA}
}
\email{mark@ucmail.uc.edu} 

\author{Ashwaq Alsoubai}
\affiliation{%
  \institution{King AbdulAziz University}
  \city{Jeddah}
  \country{Saudi Arabia}
}
\email{atalsoubai@kau.edu.sa}

\author{Jinkyung Katie Park}
\affiliation{%
  \institution{Clemson University}
  \city{Clemson}
  \state{South Carolina}
  \country{USA}
}
\email{jinkyup@clemson.edu}

\author{Pamela J. Wisniewski}
\affiliation{%
  \institution{International Computer Science Institute}
  \city{Berkeley}
  \state{California}
  \country{USA}
}
\email{pwisniewski@icsi.berkeley.edu}

\renewcommand{\shortauthors}{Ma et al.}

\begin{abstract}
  In addressing various risks on social media, the HCI community has advocated for teen-centered risk detection technologies over platform-based, parent-centered features. However, their real-world viability remains underexplored by secondary stakeholders beyond the family unit. Therefore, we present an evaluation of a teen-centered social media risk detection dashboard through online interviews with 33 online safety experts. While experts praised our dashboard's clear design for teen agency, their feedback revealed five primary tensions in implementing and sustaining such technology: objective vs. context-dependent risk definition, informing risks vs. meaningful intervention, teen empowerment vs. motivation, need for data vs. data privacy, and independence vs. sustainability. These findings motivate us to rethink ``teen-centered'' and a shift from a ``fail fast'' to a ``mature safely'' paradigm for youth safety technology innovation. We offer design implications for addressing these tensions before system deployment with teens and strategies for aligning secondary stakeholders' interests to deploy and sustain such technologies in the broader ecosystem of youth online safety.

\end{abstract}

\ccsdesc[500]{Human-centered computing~Empirical studies in HCI}

\keywords{risk detection, teen online safety, social media}

\maketitle

\section{INTRODUCTION}
Teens have increasingly recognized that social media exposes them to various online risks.\footnote{In this paper, we use ``teens'' and ``youth'' interchangeably, as prior work often uses these terms to refer to adolescents (e.g., \cite{alsoubai2025timeliness, Badillo-Urquiola2020BeyondDesign}).} Pew Research Center's latest survey in 2024 showed that nearly half of teens in the U.S. now believe social media has a mostly negative effect on their peers, a significant increase from 2022 \cite{faverio202510}. These platforms often facilitate digitally-mediated risks such as cyberbullying \cite{baldry2017school, slonje2017perceived}, sexual solicitation \cite{Hartikainen2021SafeRisks, Razi2020LetsExperiences, Dev2022FromConversations}, and harassment \cite{copp2021online, madigan2018prevalence, wisniewski2016dear}.

To address these risks, social media platforms have offered built-in features like Instagram’s Teen Accounts that detect harmful content in direct messages \cite{Instagram2024InstagramAccounts} or YouTube’s \cite{YouTube2023YourMode}, and TikTok’s \cite{TikTok2023UpdatingCouncil} ``Restricted Mode'' filters for parental control. However, researchers have long highlighted the challenges of these platform-based and parent-centered approaches: they often lack public evaluation \cite{jia2015risk} or sufficient youth input \cite{Ma2024LabelingYouTube}, leading to high false-positive rates \cite{alsoubai2024systemization}. For example, teens report dissatisfaction with Instagram's built-in reporting tools for ineffectively addressing their flagged harassment \cite{Agha2023Co-DesigningInterventions, Agha2023StrikePrevention}. \novelty{These persistent post-deployment failures exemplify a ``fail fast'' paradigm, where systemic flaws are primarily discovered after an online safety solution reaches teens and cause harm \cite{Ghosh2018SafetyControl, Wang2021ProtectionSafety, Agha2023StrikePrevention}. Our work instead explores a ``mature safely'' paradigm: proactively identifying such tensions with secondary stakeholders before deployment to vulnerable youth populations.} The HCI community has recently advocated for a teen-centered approach by \novelty{developing risk detection algorithms} based on the private social media data shared directly by youth (e.g., \cite{ali_media, razi_sliding, park2023towards, alsoubai2025timeliness}). \novelty{Moving beyond this algorithmic work to developing teen-facing risk detection systems,} \methods{we} developed a standalone system, Modus Operandi Safely (i.e., textit{MOSafely}) \cite{alsoubai2022mosafely}. Grounded in a co-design study on ethical data collection from teens~\cite{badillo2021conducting}, the system was refined through prototyping with a long-term Youth Advisory Board~\cite{ali2025teens} and a collegiate senior design team. \textit{MOSafely} is an interactive, web-based dashboard that allows teens to upload their social media messaging data and further review and evaluate risks identified by Artificial Intelligence (AI) in their messages. \textit{MOSafely} uses pre-trained algorithms to detect high-severity harms like sexual solicitation and cyberbullying~\cite{razi_sliding,ali_media,kim_cyber}, and it visualizes these findings through a card-based interface (see Section \ref{designrational}).

To ensure that any youth online safety technologies are ``mature'' and robust, HCI researchers have started to involve a wider range of stakeholders beyond the family unit in their evaluation (e.g., \cite{badillo2024towards, Caddle2024AOnline, Caddle2023DutyOnline, caddle2025building, Ekambaranathan2023HowChallenges}). Such a multi-stakeholder approach can naturally help expose tensions in implementing youth online safety technologies in the long run, due to the differing interests held by stakeholders. For example, while social service providers see the potential of AI for detecting risks in teens’ online interaction data, they worry it could violate the trust they have built with youth \cite{Caddle2023DutyOnline}. A similar tension exists between social media platforms' public claims and the user experience, where users criticized that TikTok features like screen time controls are for public relations rather than genuine risk mitigation \cite{AssociatedPress2024TikTokShows}. Furthermore, while youth online safety research often promotes self-regulation \cite{Wisniewski2017ParentalSafety, McNally2018Co-designingChildren} or a community-based approach to collaboratively mitigate online risks to youth across stakeholders \cite{Akter2022FromEquals, caddle2025building}, the social media technology industry may more favor parent-centered approaches \cite{YouTube2015YouTubeKids, TikTok2023UpdatingCouncil, Instagram2024InstagramAccounts}. Situating our MOSafely dashboard within this multi-stakeholder context, our study aims to uncover \methods{early feedback before MOSafely's deployment} and further explore the tensions in implementing and sustaining such risk detection technology for teens with \expert{three groups of} online safety experts \expert{as secondary stakeholders (as opposed to family units of teens or caregivers as primary stakeholders),} including researchers, practitioners, and industry professionals. \expert{Exploring the perspective of secondary stakeholders is as important as understanding those of primary stakeholders due to their expertise in operating the resources that determine a teen online safety solution's innovation and viability, and leading knowledge production about youth online safety by working with teens directly.} Through their feedback, we aim to answer two research questions: 

\begin{itemize}
    \item \textbf{RQ1}: \textit{What \methods{early feedback} do online safety experts have regarding \clarity{the}\textit{``MOSafely''} risk detection dashboard, \methods{before its deployment with teens?}}
    \item \textbf{RQ2}: \textit{What tensions do those experts identify in implementing and sustaining a teen-centered social media risk detection technology like our dashboard?}
\end{itemize}

To answer these questions, we conducted \methods{online} semi-structured interviews with 33 online safety experts after they viewed a video demonstration of our dashboard. We performed an inductive thematic analysis on the interview transcripts. We found that while participants praised our dashboard's clear design and its features for teen agency, they also identified weaknesses, including usability barriers in the data transfer process and a lack of clarity in how risks were presented and analyzed by the AI (RQ1). We identified five primary tensions that participants perceive in implementing and sustaining such a risk detection technology, including: the conflict between objective risk definitions and teens’ context-dependent realities; the challenge of moving from simply informing users of risk assessment to enabling meaningful intervention; the disconnect between teens' motivation and the goal of teen empowerment; the privacy dilemmas created by the need for personal data; and the struggle between an independent technology and long-term sustainability (RQ2). These findings motivate us to rethink what is ``teen-centered,'' \clarity{recognizing that for an online safety solution to be effective for teens, it must account for multiple stakeholders who operate the resources and innovation on the solution \cite{Ekambaranathan2023HowChallenges, Caddle2024AOnline, martin2023teacher}. We discuss how expert evaluation allows us to shift from a ``fail fast'' to a ``mature safely'' paradigm, identifying and addressing practical concerns before online safety solutions reach teens and conclude with implications for design} \expert{and strategies for aligning the interests of experts as secondary stakeholders to guide multi-stakeholder, responsible innovation in youth online safety.}

Our study makes the following contributions to HCI research on youth online safety: \clarity{(1) The \textit{MOSafely} dashboard, developed as a fully functional system and design probe to evaluate the concept of algorithmic risk detection with our expert participants as secondary stakeholders before engaging teens. (2) A conceptual framework of five key tensions derived from this expert evaluation, spanning the foundational conflict between objective risk definitions and teens' context-dependent realities to the practical challenges of sustainable governance. (3) A re-examination of ``teen-centered'' design, revealing that for a teen online safety technology to be viable, it must navigate the intertwined technical, social, and economic needs of its secondary stakeholders, not just the teen user alone. (4) A practical roadmap of implications for both technical design and secondary stakeholders to mitigate these tensions and allow teen-centered safety technologies to ``mature safely.'' (5) Implications for aligning experts as secondary stakeholders for youth online safety efforts}.

\section{RELATED WORK}
This section reviews prior work in two lines: \novelty{the algorithmic development} for teen-centered risk detection, and the recent trend of involving multiple stakeholders \novelty{beyond the family unit in youth online safety.}

\subsection{Teen-Centered Social Media Risk Detection}
While risk detection technologies offer promise for youth safety on social media, prior work has documented two primary challenges in their development and implementation. First, many existing technologies are parent-centered \cite{Khurana2015TheHarassment, Schiano2017ParentalOpportunity, Erickson2016TheWorld}, indicating an often counterproductive approach. Parental control apps, for example, can be privacy-invasive for teens while overwhelming parents with irrelevant information \cite{Ghosh2020CircleFamilies, Ghosh2018SafetyControl, Wang2021ProtectionSafety}, and their effectiveness often depends on parent-child communication \cite{Rutkowski2021FamilyCommunication} and assumptions of parental privilege \cite{wisniewski2024moving, wisniewski2022privacy}. 
Second, many risk detection technologies built into social media platforms performed an unfulfilled job for youth. For example, a recent study \cite{eltaher2025protecting} found that YouTube's video content moderation system failed to prevent 13-year-olds from being disproportionately exposed to harmful content within minutes of passive scrolling, compared to their 18-year-old counterparts. Oftentimes, these safety features either lack public evaluation \cite{jia2015risk} or sufficient youth input (e.g., YouTube's content labels \cite{Ma2024LabelingYouTube}), leading to high false-positive rates \cite{alsoubai2024systemization}. 

In response to these challenges, HCI researchers have advocated for a more teen-centered \novelty{approach, one that is grounded in data donated directly by youth and validated by their own lived experiences} \cite{kim_cyber, Razi_Sexual}. \novelty{Building on this, prior researchers have primarily focused on developing and validating the underlying risk detection algorithms.} For example, researchers have \novelty{analyzed media from youth-donated data to identify features of unsafe conversations} \cite{ali_media}, \novelty{developed sophisticated machine learning models to detect specific risks (e.g., vision transformers for risky images} \cite{park2023towards}; \novelty{classifiers for sexual risk experiences} \cite{razi_sliding}), \novelty{and optimized these models for real-world scenarios, such as using reinforcement learning to prioritize high-risk conversations for timely intervention} \cite{alsoubai2025timeliness}. \novelty{While such work has developed foundational risk detection algorithms, it did not provide the teen-facing systems that would operationalize them, nor the practical evaluation frameworks that can help assess their real-world deployment. An important research gap thus remains in understanding how to translate these computational models into a tangible tool that teens can, or would, actually use and control. Our work is among the first to bridge this gap by designing and, more importantly, evaluating such a teen-facing risk detection dashboard \clarity{to identify implementation tensions and allow the design concept to ``mature safely,''} \textit{MOSafely} (see more in Section \ref{designrational} \methods{about how we design and develop it with teens}).} 

\subsection{\novelty{Expanding Stakeholder Involvement in Youth Online Safety}}
\novelty{Beyond the development of risk detection algorithms,} a parallel and important line of work \novelty{has called for involving a wider range of stakeholders} beyond the family unit (i.e., teens, parents) in \novelty{evaluating and refining} youth online safety solutions (e.g., \cite{badillo2024towards, Caddle2024AOnline, Caddle2023DutyOnline, caddle2025building, Ekambaranathan2023HowChallenges}). Such a multi-stakeholder approach is important because it can help expose the tensions or trade-offs that the solutions must navigate. For example, \novelty{prior studies have interviewed social service providers} \cite{Caddle2023DutyOnline} \novelty{and caseworkers} \cite{badillo2024towards} \novelty{to understand their high-level concerns and their speculative views on ``the possibility of using artificial intelligence'' for risk detection.} While social service providers see potential for AI to detect risks in teens’ online interaction data, they also worry that this could violate the trust they have built with youth \cite{Caddle2023DutyOnline}. Similarly, \citet{Ekambaranathan2023HowChallenges} found that while children’s app developers liked the idea of an independent tool to assess apps for privacy and age-appropriateness, they preferred such tools to focus on quantitative measures, like analyzing third-party libraries and data trackers, so the responsibility of child safety can be better allocated to different experts.

\novelty{While this prior work confirms that a multi-stakeholder perspective beyond the family unit is critical for technology development and deployment for youth, its inquiry has remained largely formative by identifying high-level needs without evaluating a specific technological design.} \novelty{Our study moves beyond this formative-level inquiry.} \novelty{By grounding our expert evaluation in a tangible design probe, our \textit{MOSafely} dashboard, we uncover the implementation and sustainability tensions} that a teen-centered risk detection technology must resolve before it is mature enough for teens to use.

\section{METHODS}
This section introduces our proposed teen-centered social media risk detection dashboard design and our interview study design and data analysis.

\subsection{Proposed Teen Online Safety Solution: A Teen-centered Social Media Risk Detection Dashboard}
\label{designrational}
Motivated by the need to empower youth to navigate social media risks and actively involve them in the development of protective solutions, we developed a standalone system, Modus Operandi Safely (i.e., \textit{MOSafely})~\cite{alsoubai2022mosafely}. Grounded in a co-design study with 20 adolescents~\cite{badillo2021conducting}, we first established the dashboard's foundational requirement to function as a transparent risk-reporting tool rather than a surveillance device. The system was then refined through a year-long engagement with a Youth Advisory Board~\cite{ali2025teens}, where seven teens critically assessed early prototypes and proposed features that prioritized user agency and digital resilience. \textit{MOSafely} is an interactive, web-based dashboard that enables youth to review and evaluate risks identified by Artificial Intelligence (AI) in their private social media messages. The system allows users to upload data exported from various social media platforms, such as Instagram and X, for automated risk analysis.

\begin{figure}[h]
    \centering
    \includegraphics[width=1\linewidth]{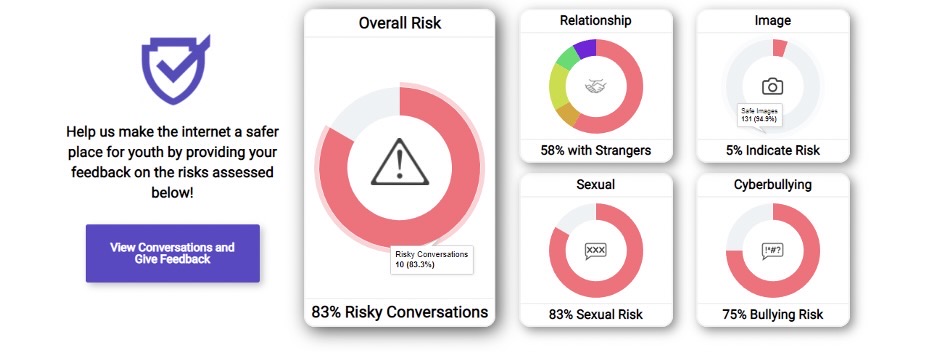}
    \caption{\clarity{The \textit{MOSafely} dashboard interface, showing card-based summaries that quantify total detected risks and break them down by category (e.g., ``Sexual,'' ``Cyberbullying'').}}
    \label{fig:dash}
\end{figure}

\begin{figure}[t]
	\centering
	\includegraphics[scale=0.2]{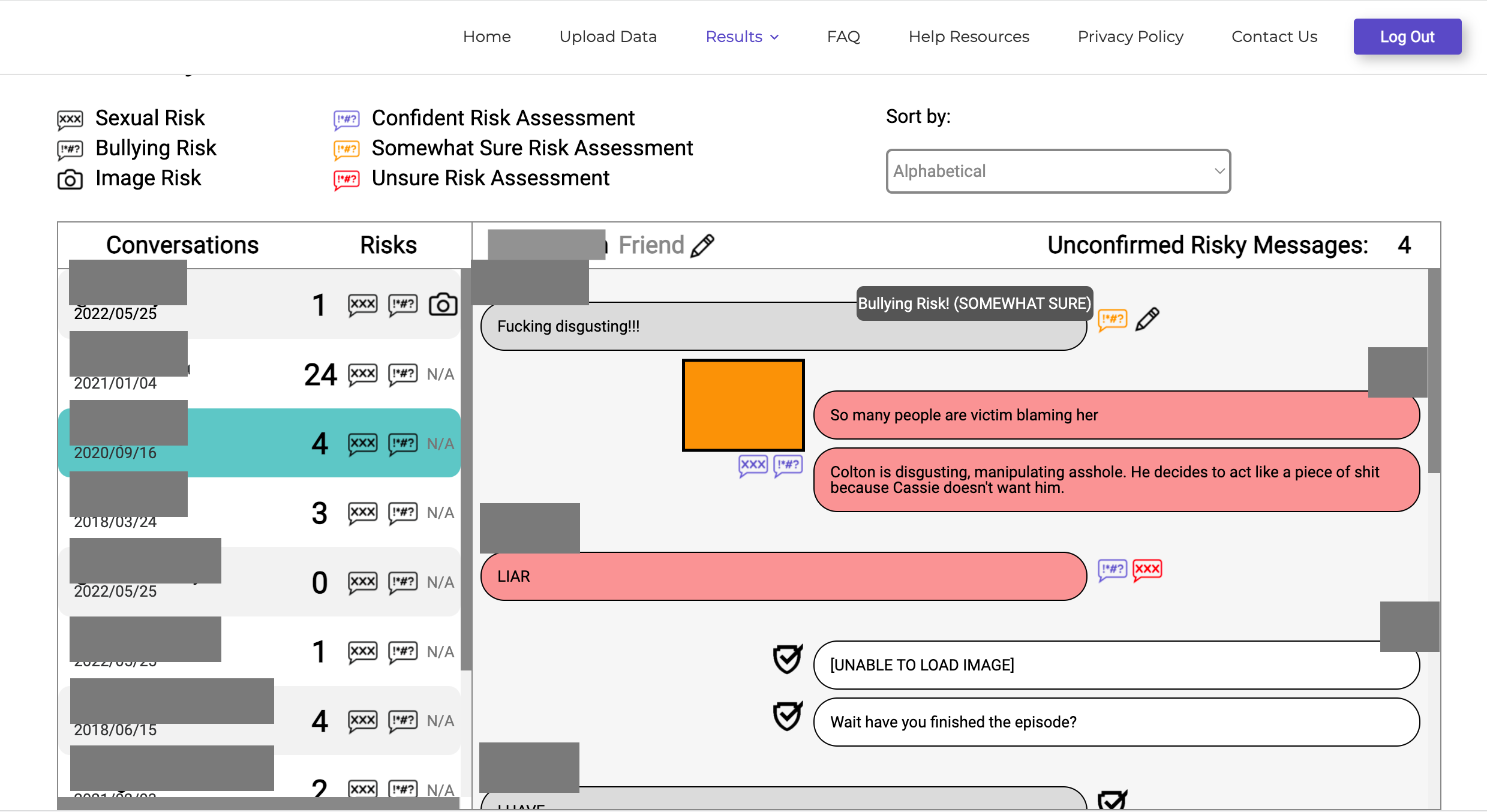}
	\caption{\clarity{The conversation management page. An edit icon (pencil) is displayed next to a message, indicating the feature that allows users to provide message-level feedback on the AI's risk assessment.}}
	\label{SMSINT}
\end{figure}

\expert{\textit{MOSafely}’s design was informed by earlier co-design and feedback sessions conducted with our Youth Advisory Board (YAB), which consisted of seven teens aged 15–17, with which we partnered for over a year. Insights from these sessions guided the development of the dashboard features and safety protocols, with teens proposing improvements that ranged from incremental refinements (e.g., clearer privacy controls) to more disruptive ideas (e.g., alternative architectures that decentralize platform control).} Based on this foundation, the system allows users to upload data exported from various social media platforms, such as Instagram and X (formerly Twitter), for automated risk analysis. Grounded in prior literature on youth and online risks (e.g., \cite{alsoubai_relationships, Hartikainen2021SafeRisks, razi2020let}), we embedded pre-trained algorithms (see Table \ref{accuracy}) within the dashboard to detect content within the private conversations associated with high-severity harms frequently experienced by youth, including sexual solicitation and cyberbullying~\cite{razi_sliding,ali_media,kim_cyber}. To provide a nuanced understanding of each flagged interaction, the dashboard also automatically detects the nature of the relationship between the youth and the conversation partner, including the stranger, friend, significant other, or family, as prior work stresses the importance of contextualizing risks by the youth's social relationships ~\cite{alsoubai_relationships}.

\begin{table}[t]
\footnotesize
\begin{tabular}{lllll}
\textbf{\begin{tabular}{@{}l@{}}Classification\\Level \end{tabular}} & \textbf{\begin{tabular}{@{}l@{}}Classification\\ Type\end{tabular}} & \textbf{\begin{tabular}{@{}l@{}}Model\\Type\end{tabular}}                                                                                   & \textbf{Accuracy} & \textbf{F1} \\ \hline
Message      & Sexual                       & DNN  & 87\%                          & 87\%              \\
                              & Cyberbullying                &                                                                                                       & 82\%                          & 82\%              \\
                              & Image                        & CNN                                                                                                   & 60\%                             & 89\%              \\ \hline
Conversation & Sexual                       & CNN                                                                                                   & 89\%                             & 90\%              \\
                              & Cyberbullying                & LSTM                                                                                                   & 68\%                             & 63\%              \\
                              & Relationship Type                 & CNN                                                                                                   & 80\%                             & 89\%              \\ \hline
\end{tabular}
\caption{\clarity{Performance of the pre-trained classifiers used in MOSafely, showing accuracy and F1 scores for both message-level and conversation-level detection. CNN denotes Convolutional Neural Networks, DNN denotes Deep Neural Network, and LSTM denotes Long Short-Term Memory.}}
\label{accuracy}
\end{table}

The dashboard visualizes detected risks through a card-based interface (Figure \ref{fig:dash}), offering a summary of the total number of risky conversations along with a breakdown by risk type and relationships detected (Figure \ref{SMSINT}). Importantly, youth are asked to assess the AI-generated risk labels and correct them, both in terms of risk presence and relational classification, and to supply additional contextual information where relevant (Figure \ref{fig:riskcorrection}). These evaluations are integrated back into the system to support iterative model refinement through a human-in-the-loop framework, promoting more accurate and context-aware risk detection over time~\cite{mosqueira2023human}. \expert{Having incorporated youth perspectives in the developmental stages, we turned to experts within the teen safety ecosystem who speak to policy considerations, platform constraints, and implementation challenges to have a holistic understanding of how such a solution will function in practice. Next, we will describe our interview procedure with those experts as secondary stakeholders to teen online safety.}

\subsection{Data Collection: Interviews with Teen Online Safety Experts \expert{as Secondary Stakeholders}} 
\label{datacollection}
Following IRB approval, we conducted \methods{online} semi-structured interviews \expert{between November 2023 and April 2024 with 33 secondary stakeholders in teen online safety. We recruited participants based on the criteria of (1) who were 18 or older, (2) spoke English, and were (3) a representative or employee of a youth-serving organization, work/worked with teens in a social or educational setting, or whose technology is used socially by teens (e.g., social media and parental control platforms). We targeted these secondary stakeholders specifically because, unlike the well-studied family unit (i.e., parents and teens), they are the professionals who are likely building youth online safety technologies, such as researchers and tech entrepreneurs, or are consulted during their development, such as educators and clinicians who work directly with teens \cite{badillo2024towards, Caddle2023DutyOnline, martin2023teacher}. Furthermore, the creation and maintenance of such technologies depend on these adult stakeholders, as they operate the socio-technical and/or financial resources \cite{Ekambaranathan2023HowChallenges, Caddle2024AOnline} that determine a online safety solution's innovation and viability. While teens are the primary end-users, we engaged this secondary stakeholder group to ``mature'' the concept and identify implementation barriers before engaging directly with vulnerable teens. Therefore, it is important to understand the underlying tensions they perceive, as their perspectives directly inform the commercial and support ecosystem for these technologies. This purposeful sampling yielded a cohort with deep experience: 10 academic researchers, 11 practitioners, such as educators, therapists, and youth advocates who work directly with teens, and 12 industry professionals in product management and technology development who lead innovation in this space. Collectively, this group possessed an average career duration of approximately 20 years, and nearly three-quarters (n=24) held advanced degrees (Master’s, PhD, or equivalent). See Table \ref{tab:participants} for their demographic information.} Participants were thanked for their time and offered a \$20 Amazon gift card, while twelve accepted.

\begin{table*}[h!]
\centering 
\scriptsize 
\caption{\clarity{Demographic and professional details of the 33 secondary stakeholders, showing their role, organization type, career field, career duration (until the date of interview), degree, and job title/description.}}
\label{tab:participants}
\begin{tabular}{lllllll}
\toprule
\textbf{ID} & \textbf{Role in this paper} & \textbf{Organization type} & \textbf{Career field} & \textbf{Career duration} & \textbf{Highest degree} & \textbf{Job title/description} \\
\midrule
1 & Researcher & Higher Education & Adolescent Health & ~26 years & PhD (2010) & Associate professor \\
2 & Practitioner & Non-Profit Organization & Health Pharmacy & ~ 23 years & Bachelor (2000) & Case manager \\
3 & Researcher & Higher Education & Computer Science / HCI & ~16 years & PhD (2015) & Assistant professor \\
4 & Practitioner & Tech Company & Clinical Psychology & ~23 years & PsyD (2001) & Therapist and Business Owner \\
5 & Industry Professional & Tech Company & Information Technology & ~25 years & MS (2005) & Business owner \\
6 & Practitioner & Non-Profit Organization & Human Trafficking Prevention & ~11 years & MSW(2015) & Youth advocate \\
7 & Researcher & Higher Education & Youth Online Safety & ~25 years & PhD (2015) & Post-Doctoral Researcher \\
8 & Industry Professional & Tech Company & Youth Online Safety & ~19 years & JD (Law); PhD (2007) & Educational software consultant \\
9 & Industry Professional & Tech Company & Educational Technology & ~27 years & MA (1995) & Content Moderation \\
10 & Practitioner & Non-Profit Organization & Human Trafficking Prevention & ~21 years & Bachelor (2003) & Youth advocate \\
11 & Industry Professional & Tech Company & Psychology / Information Technology & ~25 years & PhD (2000) & Business owner \\
12 & Practitioner & Law Enforcement & Youth Online Safety & ~15 years & N/A & Law enforcer \\
13 & Practitioner & State Government & Education & ~ 18 years & PhD (2010) & Education administrator \\
14 & Researcher & Tech Company & Psychology / Information Technology & ~19 years & PhD (2021) & Director of youth research firm \\
15 & Researcher & Higher Education & Human Development & ~23 years & EdD (2011) & Associate professor \\
16 & Practitioner & K-12 Education & K-12 Education & ~22 years & BS (2003) & Teacher \\
17 & Researcher & Higher Education & Computer Science & ~21 years & PhD (2003) & Associate Professor \\
18 & Researcher & Higher Education & Computer Science & ~5 years & PhD (2019) & Post-Doctoral Researcher \\
19 & Researcher & Higher Education & Cybersecurity & ~15 years & PhD (2014) & Associate professor \\
20 & Researcher & Higher Education & Information Privacy & ~19 years & PhD (2012) & Associate professor \\
21 & Researcher & Higher Education & Youth Online Safety & ~9 years & PhD (2022) & Assistant professor \\
22 & Industry Professional & Tech Company & Information Technology & ~14 years & Bachelor (2005) & Technical Leader \\
23 & Practitioner & Non-Profit Organization & Adolescent Health & ~19 years & DrPH (2018) & Curriculum building \\
24 & Practitioner & K-12 Education & K-12 Education & ~26 years & MS (2015) & Teacher \\
25 & Industry Professional & Independent Consulting & Product Management & ~19 years & Bachelor (1993) & Application development \\
26 & Practitioner & K-12 Education & K-12 Education & ~29 years & MEd (1994) & Teacher \\
27 & Industry Professional & Tech Company & Product Management & ~16 years & Bachelor (2009) & Application development \\
28 & Industry Professional & Tech Company & Information Technology & ~25 years & MA (2012) & Business owner \\
29 & Industry Professional & Tech Company & Adolescent Health & ~24 years & MEd (1999) & Business owner \\
30 & Practitioner & K-12 Education & K-12 Education & ~30 years & MEd (2000) & Education administrator \\
31 & Industry Professional & Tech Company & Product Management & ~20 years & Bachelor (2003) & Business owner \\
32 & Industry Professional & Tech Company & Information Technology & ~26 years & PhD (2005) & Engineer \\
33 & Industry Professional & Tech Company & Product Management & ~18 years & MA (2006) & Business owner \\
\bottomrule
\end{tabular}
\end{table*}

We structured each \methods{60-minute} online interview session, recorded and transcribed by Zoom, to leverage design probes to ground the discussion and structured it into three phases: a video demonstration of our risk detection dashboard and reflection questions (Phase 1), interaction with a community-building function (Phase 2), and a concluding discussion on broader youth safety topics (Phase 3). Although the full interview transcripts gathered data across the three phases, the analysis in this paper focuses exclusively on the feedback collected during Phase 1. \methods{This phase of the interview lasted approximately 15--30 minutes per participant.} This phase aligns directly with our research questions, which concern the \methods{early} feedback (RQ1) and implementation and sustainability tensions (RQ2) of the teen-centered risk detection technology, distinct from the broader topics explored in later phases. Phase 1 asked participants for their thoughts on the dashboard, the utility of its features, and potential barriers that prevent teens from benefiting from it. We also asked for suggestions on new features, other stakeholders who could use the dashboard, and practical considerations such as whether access should be free and how it can be sustained. We also followed up with probe questions when we encountered interesting viewpoints that required further discussion. \clarity{Phases 2 and 3 of the interview protocol involved an interactive walkthrough and subsequent discussion of an expert-facing online community concept. During this part, participants were asked to complete tasks such as creating a profile, searching for collaborators, and proposing projects on our dashboard, with the goal of fostering collaboration on research and algorithm development among secondary stakeholders. Therefore, this data was determined to be out of scope.} The dashboard shown in Phase 1 did not collect any real user data; it displayed example data created by the research team to illustrate its functionality. 
\methods{We chose this video demonstration approach because secondary stakeholders are not the primary users; requiring them to upload their adult social media data would lack ecological validity for testing a teen-centered risk detection technology. This approach allowed us to responsibly gather feedback before exposing a vulnerable youth population to it.}

\subsection{Data Analysis}
We performed an inductive thematic analysis \cite{braun2019thematic} on the interview transcripts from Phase 1. The first coder began the analysis by familiarizing themself with the dataset, reading all transcripts to assess their depth and richness. Following this, the coder started assigning initial codes to the data in Google Sheets. \methods{Alongside the thematic codes, the first coder also tagged each quotation with its corresponding participant grouping (i.e., ``Researcher,'' ``Practitioner,'' or ``Industry Professional'') based on the roles defined in Table \ref{tab:participants}.} \methods{This role-tagging was used to facilitate our later analysis of how perspectives might align with or differ across these stakeholder groups.} For example, the statement, \textit{``I think the greatest chance of like it being built in a way that is most helpful, scientifically based. trustworthy, etc. It should be a standalone and probably by a nonprofit,''} was assigned the initial code for \textit{trustworthiness of risk detection technology, it should be built/hosted by a non-profit org.''} After this initial coding, the first coder used a Miro board to affinity diagram the codes, consolidating them into sub-themes and then grouping related sub-themes into primary themes. It was during this consolidation stage that the analysis became more focused, with themes being organized to directly address our research questions on \methods{early} feedback (RQ1) and implementation and sustainability tensions (RQ2). Throughout this process, three other members of the research team regularly met with the first coder to discuss the codes, themes, and associated quotations. These sessions were used to address disagreements, refine the thematic structure, and ensure the reliability of the final analysis, which resulted in a thematic scheme to structure our findings. \methods{This approach, which involves a single primary coder who regularly meets with the research team for discussion and refinement, is well-suited for thematic analysis \cite{braun2019thematic} and a common and rigorous practice shown in prior HCI work \cite{jiang2021supporting, patel2019feel, gauthier2022will}.}

\section{FINDINGS}
\subsection{RQ1: \methods{Feedback on the MOSafely Dashboard}}
While expert participants praised our dashboard's clear design and its teen-controlled risk categorization features as empowering, they also identified barriers in transferring social media data to our standalone dashboard and a lack of clarity in how the dashboard defined and analyzed risks.

\subsubsection{\textup{\textbf{Strength: Participants Praised Clarity and Teen Agency \methods{Supported by MOSafely}}}} 
\begin{figure}[h!]
    \centering
    
    \begin{subfigure}[t]{0.4\textwidth}
        \centering
        \includegraphics[width=\linewidth]{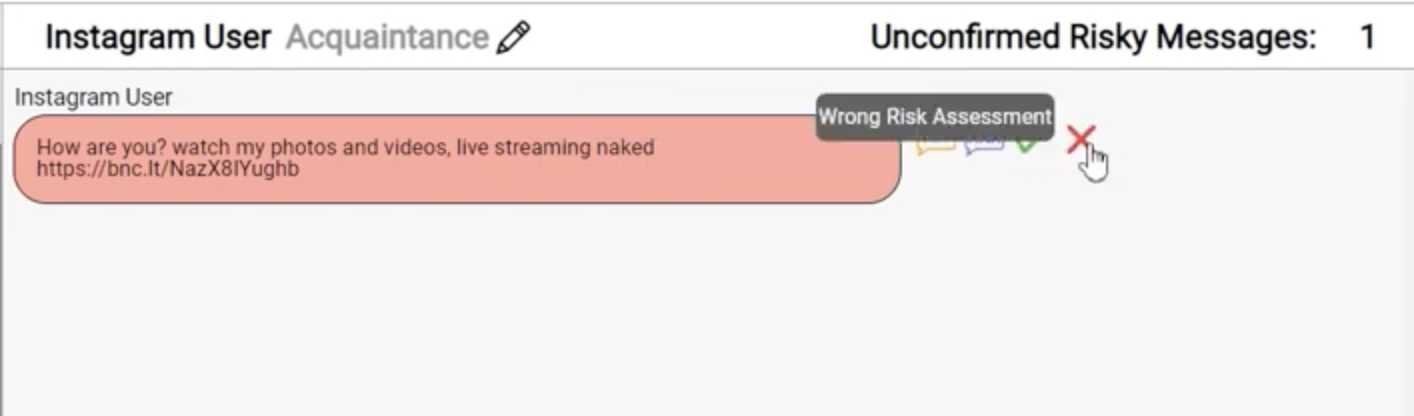}
        \caption{\clarity{A user clicks the ``Report Wrong Assessment'' button on an individual message.}}
        \label{fig:sub1}
    \end{subfigure}\quad 
    \begin{subfigure}[t]{0.2\textwidth}
        \centering
        \includegraphics[width=\linewidth]{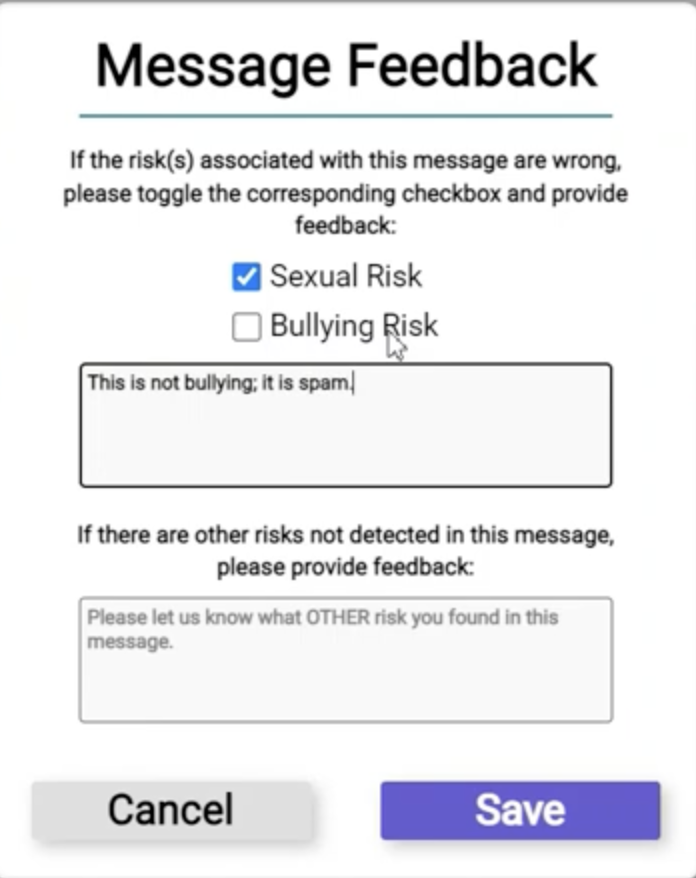}
        \caption{\clarity{A pop-up modal appears, allowing the user to select the correct risk type and provide feedback.}}
        \label{fig:sub2}
    \end{subfigure}
    
    \caption{\clarity{The two-step user workflow for correcting a misidentified risk. (a) The user reports an error on a message, and (b) a pop-up modal allows them to submit a manual correction.}}
    \label{fig:riskcorrection}
\end{figure}

\novelty{While extensive prior work has documented the prevalence of diverse online risks \cite{Faverio2024Teens, livingstone2014annual}, it also establishes that the nature of risk is highly context-dependent (e.g., differing between strangers and friends \cite{dev2022ignoring}). We found that participants endorsed our dashboard's design precisely because it addresses this known complexity.} Participants initially perceived the dashboard as easy to use, frequently citing its clean layout, clear risk categorization, and informative presentation of detection results, as P22, a technical leader, commented \textit{``I kind of like this pop-up thing to... manage the content. I think it's simple and kind of straightforward.''} Furthermore, participants especially praised the dashboard’s ability to allow teens themselves to categorize online risks at a granular level. For example, one researcher noted the usefulness of this feature in adjusting each message's risk assessment results from AI, as shown in Figure \ref{fig:riskcorrection}, while another added that it accommodates the fact that different people have different ideas about what constitutes risk.

This perspective was echoed by many participants who viewed the ability to modify risk assessment results as a form of empowerment. P26, a practitioner, remarked that with such a tool, \textit{``You don't have to be at... the hands of someone else.''} P28 further emphasized the importance of teens’ agency:

\begin{quote}
\textit{``And I like the self-selection, [which is] the agency you're giving [to] teens. That's huge. You're not being judgmental.''} [P28, industry professional]
\end{quote}

This comment illustrated how P28 saw this feature not just as a technical function, but as a mechanism that gave teens control over how their own experiences were defined, free from the judgment of an AI or others.

\subsubsection{\textup{\textbf{Weakness: Participants Identified the Data Transfer Process as a Usability Barrier}}}
\begin{figure}[h!]
    \centering
    
    \begin{subfigure}[t]{0.28\textwidth}
        \centering
        \includegraphics[width=\linewidth]{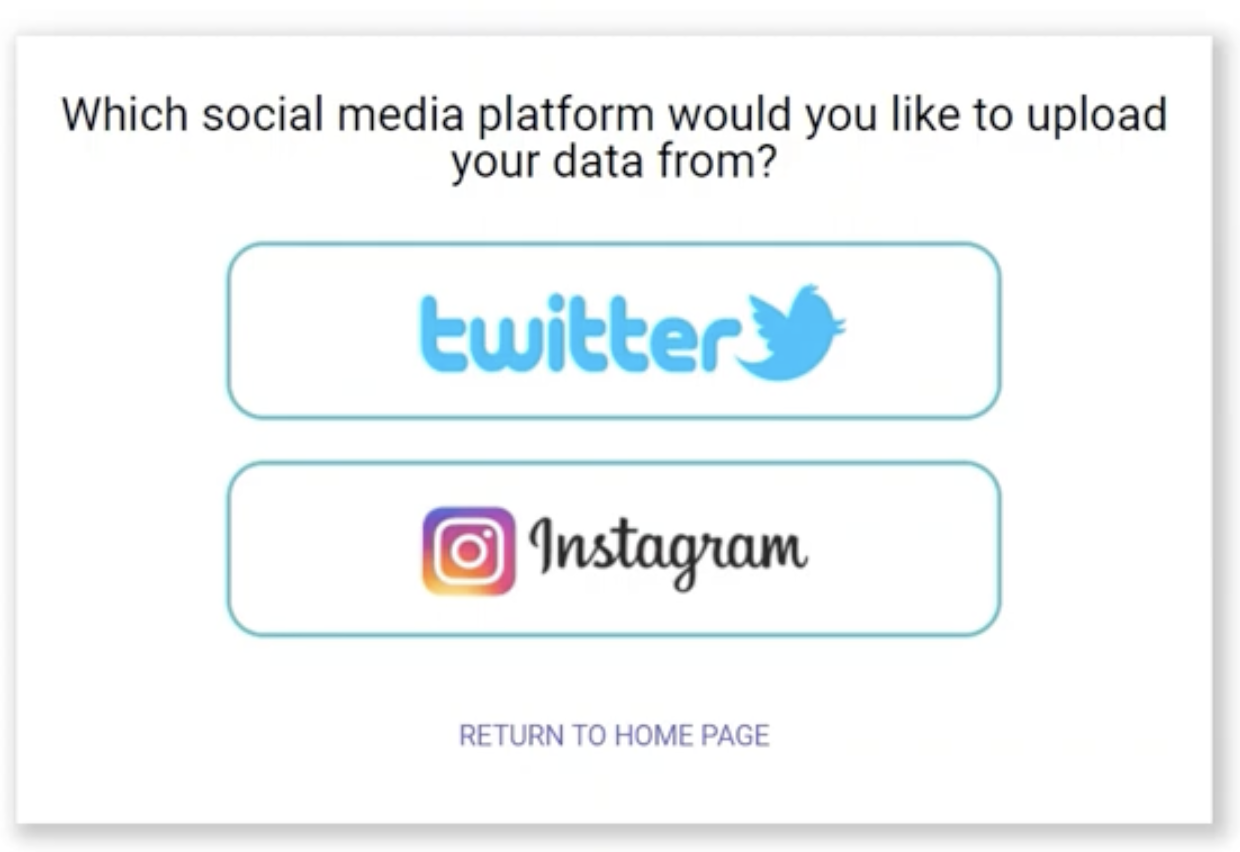}
        \caption{\clarity{The initial platform selection interface with icons for Instagram and X (formerly Twitter).}}
        \label{fig:sub_platform_select}
    \end{subfigure}\quad
    \begin{subfigure}[t]{0.3\textwidth}
        \centering
        \includegraphics[width=\linewidth]{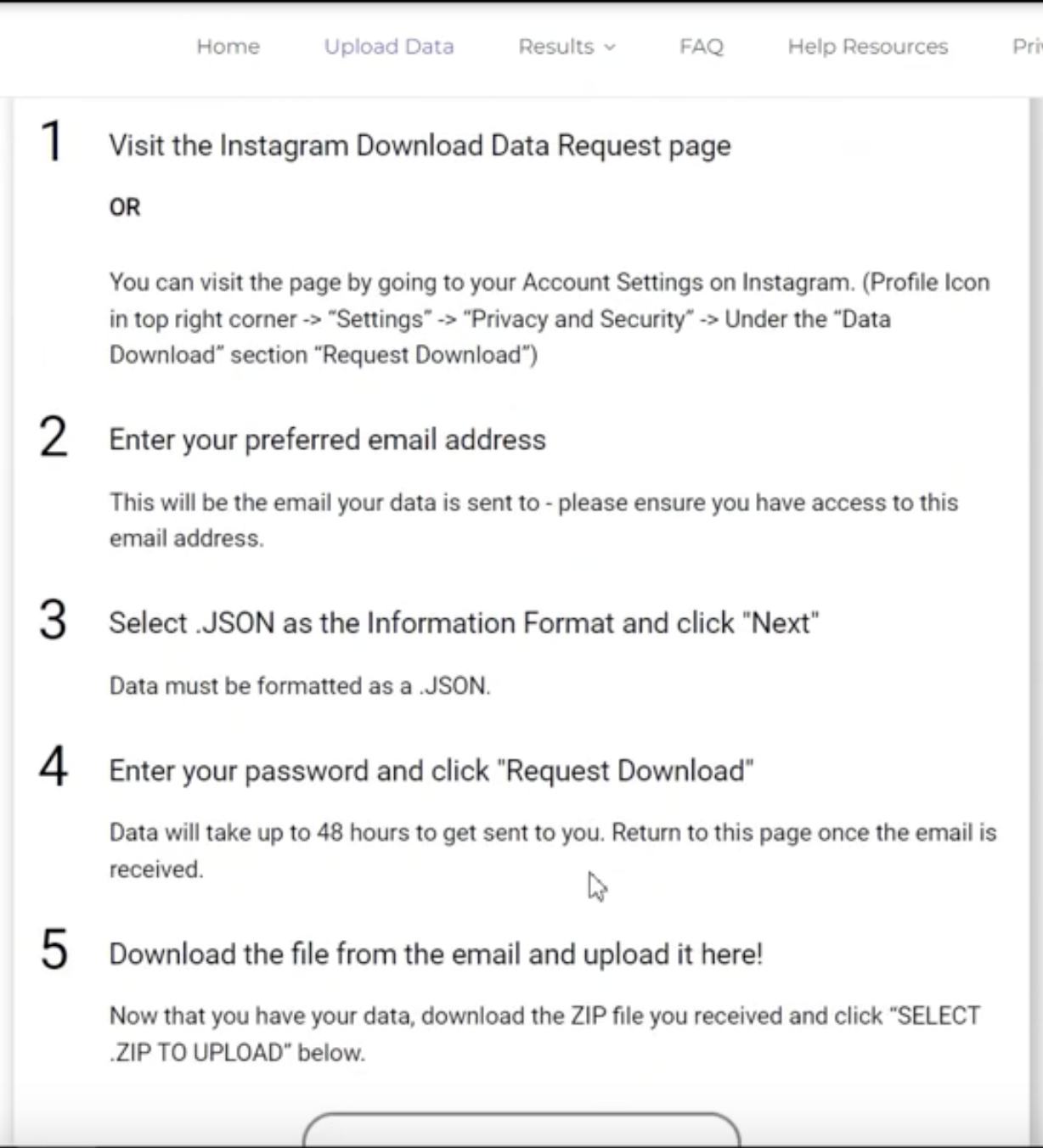}
        \caption{\clarity{A screen displaying the 5-step instructions for retrieving Instagram data.}}
        \label{fig:sub_instagram_instructions}
    \end{subfigure}
    
    \caption{\clarity{The data upload workflow. (a) The user is shown an interface to select their social media platform. (b) After selection, a new screen shows multi-step instructions for data retrieval (Instagram as an example).}}
    \label{fig:dataupload}
\end{figure}

\novelty{Prior research has documented the social and behavioral factors that protect teens online, such as their digital skills \cite{livingstone2014annual} and their own context-specific privacy practices \cite{boyd2014s, Marwick2014NetworkedMedia}. Our findings introduce novel technical barriers that undermine these protective efforts, which prior work has not previously identified.} A primary barrier mentioned by participants was that the data transfer process from social media platforms to our standalone dashboard created a usability barrier that could prevent teens’ continued use. While we provided clear instructions on how to retrieve data from platforms like Twitter and Instagram (see Figure \ref{fig:dataupload}), the lengthy instructions were perceived as a barrier by participants, as they found the instructions to be unclear and too detailed. Even when teens knew how to download their data, the process itself was seen as a hurdle. P19, a researcher, flagged that \textit{``downloading your social media data can be a bit cumbersome, and it's an effort that would be a barrier to adoption potentially.''} 

For example, P27, an application developer, articulated the multiple steps involved:

\begin{quote}
\textit{``First, teens have to want to do it themselves. Second, … there are complicated instructions on how to download your data that might change, and third, I think the frequency [of data uploading is a barrier].''} [P27, industry professional]
\end{quote}

This feedback showed that an effective data handling process for a teen-centered risk detection technology would require lightweight and clearly instructed coordination between teens, our dashboard, and the social media platforms. This concern resonated with many participants, mostly industry professionals, including P5 and P8 (business owner), and P8 (an educational software consultant).

\subsubsection{\textup{\textbf{Weakness: Participants Criticized the Lack of Clarity in How Risks Were Presented and Analyzed}}}
\label{weakness2}
\novelty{The ``black box'' problem, where AI is opaque, is a well-recognized challenge, prompting calls for greater explainability on the algorithms of online risk assessment for youth \cite{thomas2025explainable, liu2025explainable, alsoubai2023human}. Our findings contribute to this work by identifying such opacity at two practical levels, where the algorithmic logic was opaque in how risk was calculated and how the risk was conceptually defined in the first place.} First, participants noted that the meaning of risk categories was ambiguous. For instance, a researcher (P17) mentioned the need for risk definitions:

\begin{quote}
\textit{``So you're using a lot of terminology, and it is not clear if the respondents (teens) understand the nuances between different terms. So maybe if you had a question mark to say, provide help to say what cyberbullying is, what spam is, what doxing is.''} [P17, Researcher]
\end{quote}

This suggested that without clear, accessible definitions of each type of risk, perhaps through a feature like a help tooltip, it would be difficult to achieve an aligned conceptualization of risk between the dashboard and its users.

Second, participants flagged that the AI's data analysis process to come up with risk assessment results seemed opaque. This reflected a common ``black box'' problem where algorithmic tools deliver conclusions without sufficient explanation. A researcher who is an assistant professor (P21), for example, articulated that such confusion could cause:

\begin{quote}
\textit{``Image says, a hundred percent indicates risk if they click on that card... is there a way to know? What do you mean by image? cause, like image or relationship?''} [P21, Researcher]
\end{quote}

P21's comment showed how our dashboard's failure to clarify how it dimensionalizes risk (e.g., `image' vs. `relationship'), as shown in Figure \ref{fig:dash} or calculates percentages, could lead to confusion. This feedback again stresses that defining aligned risk dimensions at a conceptual level is important to preventing confusion during the risk assessment stage.

\subsection{RQ2: Tensions in Implementing and Sustaining a Teen-centered Risk Detection Technology}
\begin{figure*}
    \centering
    \includegraphics[width=0.55\linewidth]{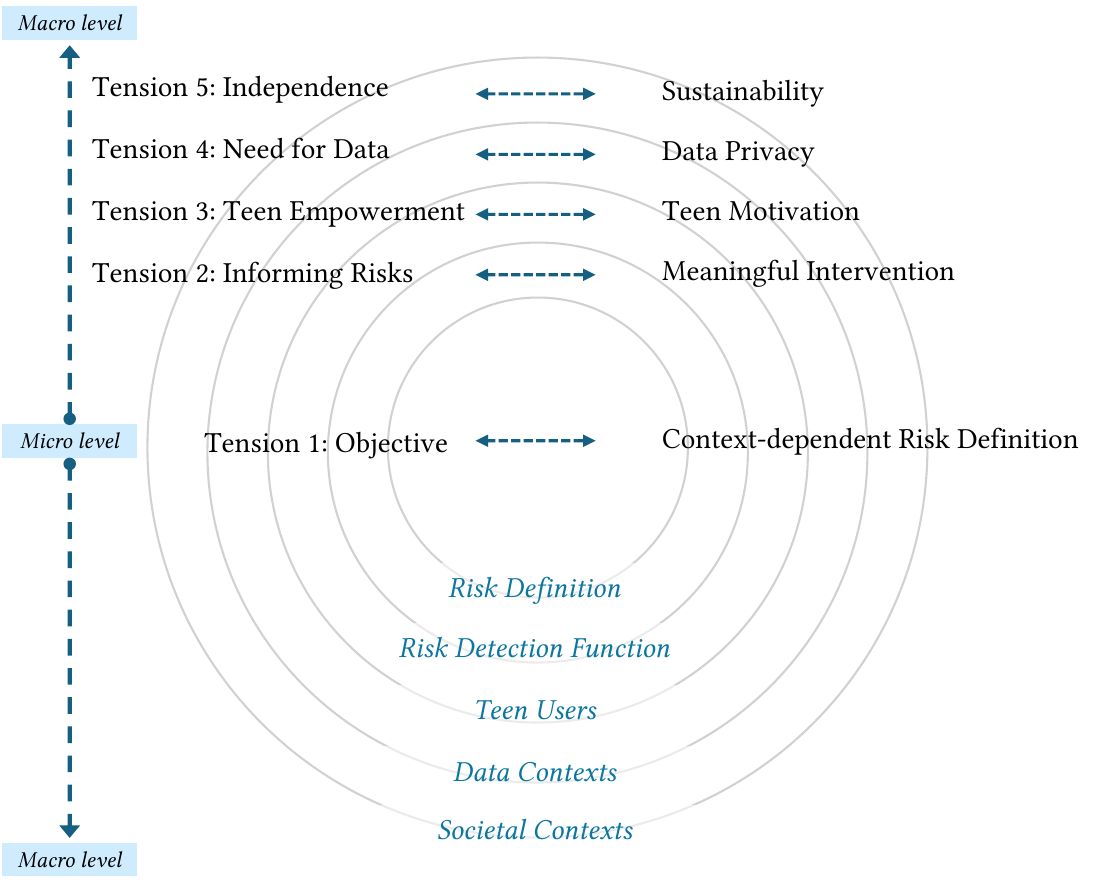}
    \caption{Five tensions that our participants identified in implementing and sustaining a teen-centered social media risk detection technology like our \textit{MOSafely} dashboard, \clarity{from the micro to the macro levels.}}
    \label{fig:rq2_visual}
\end{figure*}

Participants' feedback further revealed five tensions in implementing and sustaining teen-centered social media risk detection technologies like our dashboard. These tensions emerge across five conceptual levels (see Figure \ref{fig:rq2_visual}) and begin with the foundational conflict in Risk Definition (Tension 1), then examine the technology's Risk Detection Function (Tension 2), the Teen User it serves (Tension 3), the Data Context it operates in (Tension 4), and finally, the Societal Contexts that govern its sustainability (Tension 5).

\subsubsection{\textup{\textbf{Risk Definition: Objective Risk Definitions Often Conflict with Teens’ Context-dependent Realities}}}
\label{tension1} 
\novelty{Developmental psychology establishes that adolescent risk assessment is heavily shaped by social context, relationships, and peer norms \cite{ktoridou2012exploring, mulisa2018perceived, pinter2017adolescent, hamilton2021re}. This context-dependency is a core challenge in online spaces, which are prone to ``context collapse'' \cite{boyd2014s}. Our findings advance this understanding by documenting how expert participants perceive this as a structural design tension at the foundational level of Risk Definition, the logic of how a risk detection technology decides what constitutes a ``risk''. Participants called for more comprehensive risk categorization while acknowledging that this conflicts with the subjective, context-dependent realities of teens.} Initially, they wanted to expand what is detected, such as suicidal ideation (P23, practitioner) and scams (P7, researcher) and then expand the contextual data in risk assessment. For example, P1, who is a professor, suggested adding geo-tags to messages:

\begin{quote}
\textit{``If there's like a Geo tag…then that would be like an eye-opening experience when somebody thinks like, Oh, my God! This is a guy from California texting me, or if my kids and I saw anything outside of our neighborhood, that would be really quite alarming.''} [P1, Researcher]
\end{quote}

In this case, P1 prioritized an objective data point, geographic location, as an indicator of potential risk to teens, suggesting a desire for the technology to provide such clear-cut information.

However, participants repeatedly stressed that risk assessment must account for teens' own social contexts. They identified two areas where this tension is most apparent. First, risk detection must move beyond literal text to incorporate teens' interpretations. For example, P11, a business owner, highlighted:

\begin{quote}
\textit{``What I've heard from a lot of teens is that parents don't understand... There's a lot of stuff like language that's being used like, 'Oh, my God, I just wanna die,' right? Well, that could actually mean someone is thinking of killing themselves, or it's just the way some kids talk. And it actually has no safety concerns at all.''} [P11, industry professional]
\end{quote}

P11’s comment here suggested that teens might view the risk detection technology as an adult-driven tool that fails to understand their nuanced communication, thereby questioning its accuracy. To counter this, our participants stressed the need to center the teen’s voice. As P28, another business owner, argued, the language used by the tool must \textit{``come right directly from your target audience, [who are teens], preteens, young adults.''}

Furthermore, participants noted that the complexity of teens’ relationships with others challenges objective classification of risks. P23, an practitioner, questioned the clarity of relationship labels from a teen’s perspective:

\begin{quote}
\textit{``Will youth know the difference between an acquaintance and a friend? And how is it just like, on good faith that they are honestly picking the right drop-down item?''} [P23, practitioner]
\end{quote}

This highlighted a challenge that extends beyond our dashboard's design. The concern is not the accuracy of its risk definition, but whether teens can consistently apply these objective labels to their own complex and subjective social lives. Therefore, even with transparent risk definitions intended to help, the interaction becomes difficult when a teen is uncertain about the nature of a relationship. This concern was echoed by P29 (business owner):

\begin{quote}
\textit{``…so they might then have something that was a little bit risky, but they went and reclassified it, and said, no, this wasn't bullying. This was just my friend teasing me or something …So I was worried that some youth may not be able to be as objective.''} [P29, industry professional]
\end{quote}

P29’s worry depicted the tension: a teen’s subjective assessment, shaped by their personal relationship, may directly contradict an objective risk classification and lead them to downplay a true risk.


\subsubsection{\textup{\textbf{Risk Detection Function: A Tension Exists Between Simply Informing Users of Risk Assessment Results and Enabling Meaningful Intervention.}}}
\label{tension2}
\novelty{HCI research has co-designed online safety solutions with youth \cite{McNally2018Co-designingChildren, ashktorab2016designing} like in-the-moment nudges \cite{Agha2023StrikePrevention, Agha2023Co-DesigningInterventions} to address various risk scenarios \cite{Dev2022FromConversations}. Our findings contribute an expert-driven perspective, where participants identified a tension at the Risk Detection Function level, between the risk detection technology’s function as a passive informational platform and post-detection intervention among teen users.}

The obvious goal of a risk detection technology was to present a clear risk assessment result. As P31, a business owner, stated, it should \textit{``show me an assessment level, a chart, a percentage…of sort of the risk status of an interaction.''}. This reflected a need to distill social media message data into clear statistics that users can easily understand. Building on this, another industry professional, P33, added that risk detection technology like our dashboard should provide teens \textit{``actual value like, relationship X 22\% with strangers''}. This level of detail provides clear and actionable insight into risk source, which in turn can help users better understand and trust the overall risk assessment.

However, many participants stressed that simply providing an assessment was insufficient; they called for real-world, post-detection strategies. They wanted to know, as P12 (law enforcer) put it, \textit{``What's the... next step after you've got this dialed in?''} Participants identified two primary intervention strategies that a risk detection technology should support.

The first strategy was to integrate educational resources to mitigate harm. In detail, P21, an assistant professor, suggested providing \textit{``a small tutorial to maybe tell them (teens) more about certain risky behaviors''} to help them better understand the nature of those risks. Going beyond definitions, P31 (business owner) added that the dashboard could provide examples of how to address a detected risk. P31 explained that if teens \textit{``can actually look at something that was flagged as risky, and see what risky looks like as assessed by this tool... they can probably learn something from the assessment.''} This feedback showed that participants felt the technology should provide educational resources after detecting a risk, which could serve as a reference for teens to inform their own further risk mitigation practices.

The second strategy was to offer real-time or proactive safety nudges. This idea was mostly raised by industry professionals who considered how the dashboard might integrate with existing social media platforms. For example, P8 suggested the tool could be powerful if it helped screen messages before they are opened:

\begin{quote}
\textit{``This could be really useful if it were people that you don't know, and you're screening it before you open messages on Instagram or Twitter, so that we can know what kind of content is, and if there's a risk for you.''} [P8, industry professional]
\end{quote}

P8’s comment stressed a proactive approach that alerts teens to potentially risky messages, which can empower them in their decision to engage with such messages. P31 further added that after being alerted, teens could then utilize platform-native content moderation features, such as reporting, to flag the messages to the platforms.


\subsubsection{\textup{\textbf{Teen Users: A Tension Exists Between the Risk Detection Technologies’ Teen Empowerment Goal and the Perceived Realities of Teens' Motivation}}} \label{tension3}
\novelty{While recent participatory design research advocates for empowering teens as co-creators or youth advisors \cite{ali2025teens, noh2025youth, Anyon2018Youth-LedPrograms.}, our findings also highlight that motivational barriers like social stigma persist \cite{ali2025teens}. We help further document how expert stakeholders perceive this as an implementation tension at the Teen User level: they saw a direct conflict between the ideal of teen empowerment and the perceived motivation of teens as the most viable or motivated primary users.}

Initially, our participants argued that to be truly empowering, a teen-centered technology must provide both internal and external resources. On the internal side, the risk detection technology should offer insights that help a teen better understand themself. For example, P20, who is an associate professor, envisioned a state where teens \textit{``get very detailed output…[to] figure out what they actually think would be useful in terms of uploading my data.''} P20’s comment suggested that a key condition for teens to engage with the risk detection technology, like our dashboard, was whether it provided detailed feedback on their own thinking and behaviors, thus helping them achieve self-understanding. On the external side, participants mentioned the importance of a broader community support system for teens. P6, a youth advocate, suggested that such a technology should find \textit{``a way to really empower them (teens) and say your insight is really valuable like we're learning from you through this.''} This comment shows that P6 believed teen empowerment could also come from positive feedback from adults who value the insights enabled by teens' data.

However, participants questioned the viability of teens as the primary users of such technologies, citing three reasons that challenge the empowerment goal. First, many participants, including researchers P15 and P20, who are both associate professors, and practitioners P2 (case manager), P10 (youth advocate), P12 (law enforcer), and industry professionals P8 (educational software consultant), P33 (business owner), perceived that teens lack strong intrinsic motivation for online safety tasks. P2 commented, \textit{``if the kids have to upload it, I don't know necessarily how willing they would have been to do that themselves.''} P12 elaborated on this perceived lack of motivation, stating that teens would not act \textit{``unless there's a kind of goal if they do that.''}

Second, participants suggested that teens might not adopt the technology due to social stigma. P13, an education administrator, initially commented that \textit{``they (teens) wouldn't want to feel embarrassed or ashamed of engaging in this process.''} P30, another education administrator, articulated this stigma in detail, explaining that broad adoption would require community-level participation to make it socially acceptable:

\begin{quote}
\textit{``So this would have to be something that a network of parents or a school community, or a religious community participates in to lower the stigma... They won't do it because the fear of being stigmatized as a looper, a geek, a nerd, whatever is so high, there is nothing, not money, nothing that will overcome that.''} [P30, practitioner]
\end{quote}

In P30’s view, the social stigma of using an online safety tool, being seen as different or uncool, presents a barrier, reflecting the challenge of how perceived social norms affect the adoption of youth-centered risk detection technologies.

Finally, participants saw parental involvement as a key aspect of fostering community-wide adoption of risk detection technology like our dashboard. A postdoc researcher (P7) noted that parents are highly motivated to use such technology, suggesting they might require their teens to use it as a condition for joining social media platforms. While this top-down approach could be perceived as restrictive, participants suggested it would be effective if integrated into a larger community or school-based initiative, thereby normalizing its use and reducing the social stigma for teens.





\subsubsection{{\textup{\textbf{Data Contexts: Risk Detection Technology's Need for Personal Data Creates a Privacy Dilemma.}}}}
\label{tension4} \novelty{Prior work has documented the legal and ethical barriers to collecting data from minors in research \cite{alsoubai2023human, canosa2018reflexivity, ali2025teens}. Our findings add a new dimension to this work by documenting the tension at the Data Context level, which concerns the information that a risk detection technology needs versus its privacy implications. Expert participants initially affirmed the value of this data, believing that risk assessment could be used to augment teens' long-term risk awareness.} P15, a researcher, noted that with this data, teens could \textit{``get a sense of over time of what their data looks like,''} highlighting the value of seeing temporal risk patterns. Participants also felt the technology’s outcome should be to help teens understand risks they might have previously underestimated. As P6, a youth advocate, explained:

\begin{quote}
\textit{``(If kids are saying…) ‘No, this is the person I'm talking with, so this isn't a risk,’ when actually it still might include some kind of coercion or extortion. If the AI could pick up on very real, legitimate risks that kids face, [that would be helpful.]''} [P6, practitioner]
\end{quote}

In P6’s comment, the risk detection technology’s role was not only to deliver risk assessment results, but to effectively use those results to augment a teen’s risk awareness, potentially changing their perception of risk severity over time.

However, the need for teens’ personal data elicited major privacy concerns. To achieve this enhanced awareness, risk detection technology like our dashboard must first access teens' personal messaging data, which is the step that many participants found deeply problematic. They raised privacy concerns from three perspectives.

First, they projected that teens would distrust the process of data donation. This concern was flagged by researchers and industry professionals alike, P21 (assistant professor) commented, \textit{``I'm not sure how comfortable they (teen) would feel about...uploading their private data,''} and P9 (content moderation professional) similarly noted, \textit{``that's a question that's I would expect anybody to want to know.''} Experts worried this distrust might outweigh the tool's benefits, especially for the most vulnerable. P6, a youth advocate, explained:

\begin{quote}
\textit{``But I think for some kids like a barrier might be that they are actually in really high-risk situations. That could create Difficulty in wanting to share their personal messages because deep down inside, maybe they know there's a lot of stuff in there that's not OK, but they aren't ready to share that.''} [P6, practitioner]
\end{quote}

In P6’s comment, this perceived teens' distrust was rooted in the suspicion about who might see their data, an uncertainty of whether the risk assessment results would be visible only to the teen or to others. This more speaks to an interpersonal privacy concern. 

Second, participants questioned the technology's adherence to legal and regional privacy regulations. P19, a researcher in the privacy and security domain, noted that data collection from teens would need to go through an IRB so \textit{``everything is protected.''} P33, a business owner, pointed to international law, stating, \textit{``You have to ask Instagram to get a download of a JSON file...cause that's part of the GDPR,''} framing the process as a significant hurdle. 

Participants also raised regional-level privacy concerns, as P14 commented, \textit{``data privacy concerns vary by state.''} In detail, P3, an assistant professor, used an analogy to illustrate the legal complexity of data ownership across regions:

\begin{quote}
\textit{``There's a lot of states for recording your phone conversations, legal or illegal... social media is weird like that. It's never your data, right?''} [P3, researcher]
\end{quote}

P3’s analogy implied that data ownership is not exclusive to the teen but can be shared with others, including the social media platform itself, creating commercial privacy concerns for any risk detection technologies to clarify.

Third, participants warned that handling data from unconsenting third parties is an ethical concern. Because messages inherently involve more than one person, the risk detection technology, like our dashboard, would naturally analyze data from individuals who might never have consented. As P20, a researcher, stated, \textit{``Messages are by their very nature between 2 or more people.''} This led P22, a technical leader, to flag the ethical implications:

\begin{quote}
\textit{``That you are downloading actual messages to that user, so it's other people's data that you're actually analyzing. So I don't know the legal implications around that, but that's something you might want to consider.''} [P22, industry professional]
\end{quote}

P22’s comment highlighted the ethical question of whether a teen-centered risk detection technology can operate without acquiring consent from every person whose data is being analyzed.

\subsubsection{{\textup{\textbf{Societal Contexts: The Ideal of Independence of Risk Detection Technology Struggles Against Its Sustainability.}}}}
\label{tension5}

\novelty{Prior work on youth-serving nonprofits has documented the financial vulnerabilities \cite{ouellette2020systematic} and funding inequities they face \cite{gooden2018examining}. Our findings extend this organizational context to a technological one, showing an important tension at the Societal Context level: participants championed an independent, non-profit model of risk detection technology for trustworthiness but warned that its financial model would create a direct trade-off with the equity and representativeness of its user base.} Initially, participants, including P11 (a business owner), argued that the \textit{``greatest chance of it being built in a way that is most helpful, scientifically based, and trustworthy... is that it should be a standalone and probably by a nonprofit.''} This comment underscored a belief in the importance of technological independence from large social media platforms to protect the best interests of teens.

However, participants also identified two primary barriers to sustaining such a teen-centered risk detection technology. First, they were concerned about finding a sustainable financial model. P14, a researcher, noted, \textit{``it's very hard to get them funded other than through for-profit funding mechanisms,''}. P11 (business owner) further elaborated:

\begin{quote}
\textit{``if it's not free, then you definitely won't get teens paying for it on their own. And so then you would have to rely on parents or schools. And then, if you rely on schools. It's gonna be the ones that have the most funding... the data that's entered in is not gonna be fully representative. It'll start to be skewed or affluent teens.''} [P11, industry professional]
\end{quote}

As P11 reflected, a tension existed between the financial sustainability of the technology and the goal of achieving representative risk assessment, as a pay model could skew the user base toward more affluent communities.

Second, some participants warned that the technology's sustainability depends on anticipating and governing misuse by its users. An industry professional (P5), for example, worried:

\begin{quote}
\textit{``if you cared enough, you could attempt to use a generative AI to take your original messages... take that same phrase to generative AI to tone it down a bit and then put it back in there so someone can masquerade or hide themselves.''} [P5, industry professional]
\end{quote}

P5’s comment showed that sustaining the risk detection technology requires ongoing maintenance to ensure policy-compliant use, given that some users might try to deceive the tool and invalidate the risk assessment. Similarly, participants worried that teens would simply find ways to avoid the technology altogether. P30, an practitioner, used their stepdaughter as an example of this active resistance:

\begin{quote}
\textit{``I have a stepdaughter who's 14… She will do everything she can to get around any safeguards we try to put in place... She actively seeks out the risk of social media.''} [P30, practitioner]
\end{quote}

While P30 was generalizing from a personal experience, their comment reflects a broader concern about how teens' natural risk-taking and boundary-testing behaviors must be factored into a technology's design to manage avoidance for its long-term sustainability.


\section{DISCUSSION}
Our evaluation with experts \expert{as secondary stakeholders} of a teen-centered risk detection dashboard reveals findings at two levels. First, regarding design, experts praised the dashboard's clarity and support for teen agency but identified immediate weaknesses in data transfer and risk presentation (RQ1). Second, their feedback exposed five deeper tensions in implementing and sustaining such technology, ranging from objective vs. contextual risk definitions to challenges of motivation, privacy, and sustainability (RQ2). This feedback does not merely reflect usability issues but symptoms of underlying tensions. To unpack this, we re-examine what ``teen-centered'' means within a multi-stakeholder context and articulate how the expert perspective provides a practical model for responsible innovation for teen online safety.

\subsection{Rethinking ``Teen-Centered'' for Online Risk Detection}
The HCI community has long advocated for a ``teen-centered'' approach, prioritizing youth agency through methods like co-design and research apprenticeships to move beyond restrictive, parent-centered models \cite{ashktorab2016designing, badillo2019stranger, Agha2023StrikePrevention, chatlani2023teen} in online safety solutions. However, our expert participants questioned whether a singular focus on the teen user \expert{or parents} is sufficient for real-world viability. \expert{They as secondary stakeholders may operate certain technical and financial resources that influence the development of an online safety solution, as well as the knowledge production that guides the broader safety ecosystem that supports youth. Consequently, }our findings reveal that teen-centered technology must navigate a network of \novelty{multiple} adult stakeholders to be sustainable, necessitating a rethinking of ``teen-centered.''

First, we identify a technical misalignment in risk definition, where adult-led design can privilege adult knowledge over teen experience \cite{poole2013interaction, davis2020co}. Our participants called for objective indicators (e.g., scams, geo-tags) while simultaneously admitting that teens' nuanced language and complex relationships often defy such classification (Section \ref{tension1}). This reflects a broader tension in safety research where AI algorithms, often trained on adult-defined labels \cite{Razi_Sexual}, misinterpret activities like sexting that teens may view as relationship norms rather than crimes \cite{gewirtz2018complex, razi2020let}. Consequently, algorithms may misread teen communication, perpetuating the very power imbalances teen-centered technology aims to solve.

Second, our findings challenge the notion of the ``teen user'' socially, revealing that designing \textit{for} teens does not guarantee use \textit{by} teens. Echoing research on technology non-use \cite{satchell2009beyond, baumer2015study}, expert participants' doubts about teen motivation reflect a form of intentional non-use driven by ``disinterest'' in adult-prioritized topics or fear of social stigma. This creates a dilemma contrasting with typical user-centered design, where the user and ``customer'' align.

Third, the ideal of independent risk detection technologies serving teens' best interests confronts the political-economic realities. While participants championed a non-profit model as most trustworthy, they warned that financial sustainability might necessitate for-profit models or reliance on institutions like schools. This exposes a conflict between safety values and commercial business models that prioritize user engagement \cite{docherty2022re}. Furthermore, the labor required to maintain interoperability with closed social media platforms, exemplified by the cumbersome data transfer process, adds reflexive nuance to prior co-design work \cite{Agha2023StrikePrevention, chatlani2023teen}: without significant resources to overcome these political-economic barriers, such solutions may struggle to be actionable or sustainable.

These technical, social, and political-economic misalignments are not discrete but are intertwined. A technical decision about how to define risk has social consequences for whether teens trust the tool, while a social need for adoption is constrained by the financial reality of who pays. This interdependence requires the HCI community to adopt a systemic view when designing \textit{for} or \textit{with} teens (e.g., \cite{Wang2023TreatOnline, McNally2018Co-designingChildren, Agha2023StrikePrevention}) that must account for the often invisible work, such as the legal compliance, financial support, and data management, that sustains any youth online safety solution. Adopting this definition of ``teen-centered'' is therefore an important step for future teen online safety.

\subsection{The Role of Experts in Responsible Innovation for Teen Online Safety Solutions}
Prior work has well documented the persistent failure of deployed youth safety features, from the low adoption of parental controls \cite{Ghosh2018SafetyControl, Wang2021ProtectionSafety} to the ineffectiveness of platform-based tools \cite{Agha2023StrikePrevention}. Our study extends this work by using expert evaluation to proactively identify the root causes of these failures. The early feedback, such as cumbersome data transfer, is not merely a usability issue but a symptom of systemic tensions that could cause a standalone technology to fail. This suggests that the ``fail fast'' approach, waiting to discover flaws after reaching teens, is too late; instead, a responsible approach must identify and resolve these issues early in the design lifecycle for online safety solutions.

Echoing the recent trend in HCI to involve more stakeholders beyond the family unit in youth online safety solution design and evaluation (e.g., \cite{Badillo-Urquiola2024TowardsCare, Caddle2023DutyOnline, caddle2025building}), we synthesize these expert perspectives into three dimensions for maturing risk detection technology: ethical viability, real-world impact, and long-term sustainability (Table \ref{discussiontable}). For example, regarding ethical viability, researchers prioritized data ethics, while industry professionals emphasized compliance (e.g., GDPR). Regarding real-world impact, practitioners focused on connecting teens to support, whereas industry professionals focused on scalability. This mapping reveals two primary ways experts contribute to innovation.

First, these diverse experts serve as a checkpoint for \textit{responsible innovation}, anticipating and addressing the ethical or societal implications of new technologies under uncertainty \cite{stilgoe2020s, bates2019towards}. Our participants engaged in necessary speculation, identifying challenges like GDPR compliance (Section \ref{tension4}) and non-profit financial viability (Section \ref{tension5}), which are not typically surfaced in prior teens' online safety work (e.g., \cite{chatlani2023teen, Akter2022FromEquals, Wisniewski2017ParentalSafety}). Identifying these challenges before technological deployment ensures that by the time an online safety solution reaches teens for summative evaluation \cite{pinter2017adolescent}, it is already robust and responsible.

Second, experts act as proxies, helping surface competing values. While prior work effectively measured value tensions within the family (e.g., safety vs. autonomy \cite{czeskis2010parenting, Ghosh2020CircleFamilies, nouwen2018towards}), our study reveals tensions among secondary stakeholders. For example, critiques of opaque AI (Section \ref{weakness2}) highlight the tension between technical goals and \textit{transparency}, while financial concerns (Section \ref{tension5}) reveal conflicts between \textit{sustainability} and \textit{fairness} or \textit{equity}. This demonstrates how multi-stakeholder evaluation reveals competing values arising from diverse interests.

\begin{table*}[t]
    \centering
    \caption{\clarity{A summary of secondary stakeholder perspectives on maturing a teen-centered risk detection technology, categorized by challenge and stakeholder role.}}
    \label{discussiontable}
    \includegraphics[width=0.7\linewidth]{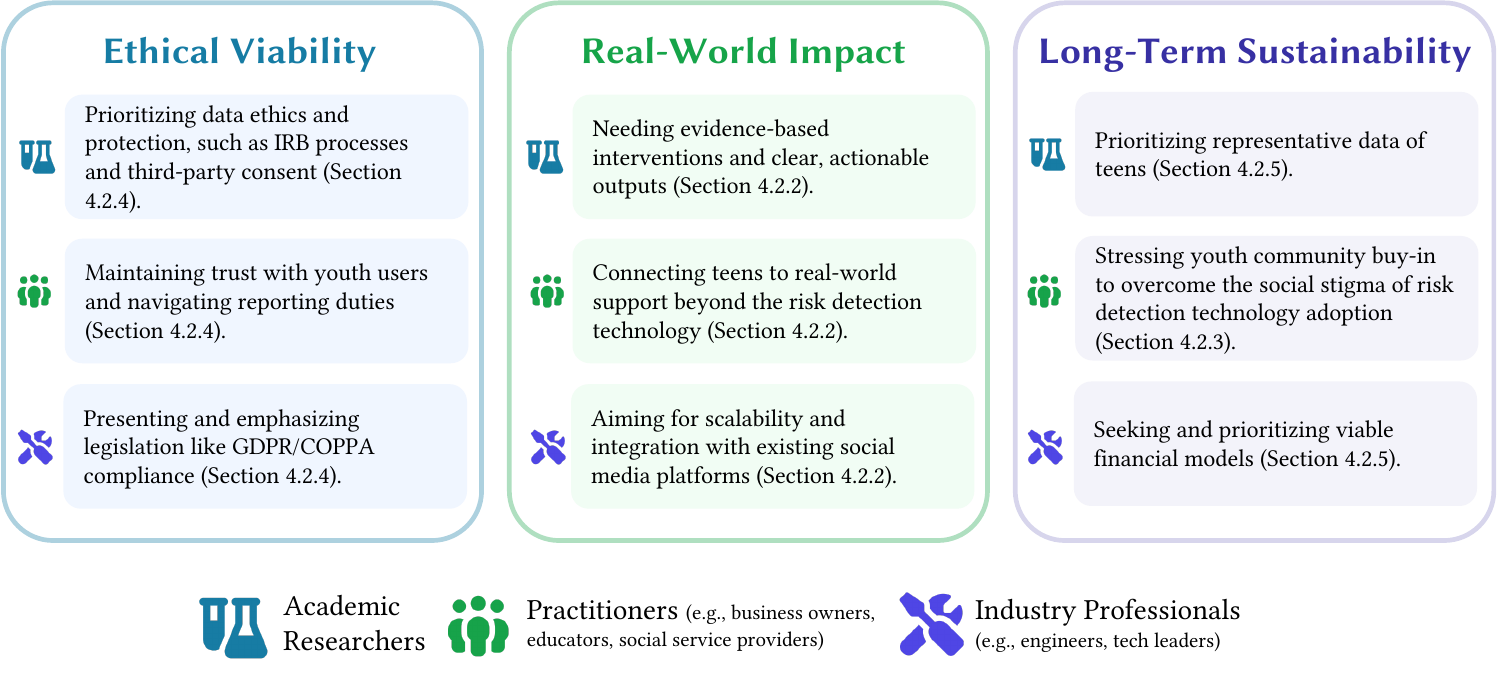}
\end{table*}

As principles like \textit{Safety by Design} \cite{badillo2019stranger, Ghosh2020CircleFamilies} call for building safety into the core of online platforms by empowering youth with proactive and self-regulation tools, our study demonstrates how a secondary-stakeholder, expert-informed layer is also important for this process. By moving beyond identifying singular value tensions (e.g., safety vs. autonomy \cite{Wisniewski2017ParentalSafety}) to mapping the competing stakeholder values, we need to stress-test the broad sociotechnical system where a youth online safety solution must exist. Thus, our practice of early-stage system development through \textit{MOSafely} is not merely to validate a user-facing concept, but to proactively address the deep-seated tensions that determine a technology's real-world success, ensuring that the solutions we build are not only effective but also sustainable.

\subsection{Implications for Design: A Roadmap for Mitigating Tensions in Teen-Centered Risk Detection}
Our findings provide a practical roadmap for maturing teen-centered social media risk detection technologies, where the tensions identified by our participants can be translated into design implications that address these challenges.

\textbf{From Objective vs. Context-dependent Risk Definition $\to$ Design for Hybrid Intelligence with Reflexive Feedback.} To resolve the conflict between objective risk definitions and teens' lived experiences (Section \ref{tension1}), risk detection technology should leverage hybrid intelligence where teen feedback continuously refines it \cite{mosqueira2023human, ali_media}. Rather than treating user input as a mere error correction, it can highlight ambiguous cases (e.g., \textit{``Is this teasing or bullying?''}) and solicit context through lightweight, conversational prompts. This approach respects the teen's interpretation as the ground truth while generating the labeled data necessary to align the AI with their specific social context \cite{kim_cyber}.

\textbf{From Informing Risks vs. Meaningful Intervention $\to$ Design Tiered and Context-Sensitive Support.} To bridge the gap between passive information and active intervention (Section \ref{tension2}), designers should implement adaptive escalation strategies \cite{Agha2023Co-DesigningInterventions} by offering a spectrum of supports: minor risks might trigger reflective self-regulation prompts, while severe risks provide direct links to real-time crisis lines or platform reporting tools. Crucially, these intervention tiers should be customizable, allowing teens to pre-select the level of support they find helpful rather than intrusive, thereby preserving their autonomy while ensuring safety \cite{Wisniewski2017ParentalSafety}.

\textbf{From Teen Empowerment vs. Motivation $\to$ Design for Scaffolded Agency with Social Reinforcement.} To counter the tension between empowerment goals and low user motivation (Section \ref{tension3}), risk detection technology must scaffold agency through social reinforcement rather than relying solely on intrinsic safety motivations \cite{hamilton2021re}. Strategies could include an anonymized community dashboards that visualize aggregated risk patterns (e.g., \textit{``50 peers flagged similar scams this week''}) to normalize the safety behavior and reduce stigma. By framing risk detection as a collective effort rather than an individual burden, designers can leverage peer influence to sustain engagement \cite{badillo2019stranger}.

\textbf{From Data Needs vs. Privacy $\to$ Design for Transparent, Time-Limited Data Practices.} To navigate the dilemma between algorithmic data needs and privacy distrust (Section \ref{tension4}), risk detection technology should prioritize transparent processing \cite{wisniewski2022privacy}. \novelty{Given the ethical gravity of granting an AI tool access to a teenager's intimate conversations and social circles \cite{o2025school, yu2025understanding}, technical efficiency cannot supersede privacy preservation. We argue for a standard of data minimization by default, where risk detection technology can access only the specific message segments required for immediate analysis, rather than ingesting entire conversation histories, which poses significant surveillance risks.} Consequently, the technology should offer ``scan-and-delete'' workflows where data is analyzed locally without permanent storage, accompanied by visual notices explicitly stating when data is purged. Furthermore, to address ethical concerns regarding third-party data \novelty{and the potential for non-consensual monitoring}, developers and designers should employ privacy-preserving techniques (e.g., federated learning or synthetic data augmentation) that minimize the exposure of non-consenting conversation partners \cite{Ghosh2018SafetyControl}.

\textbf{From Independence vs. Sustainability $\to$ Design for Modular Integration with Ethical Guardrails.}To balance the need for independence with long-term sustainability (Section \ref{tension5}), the technology should be architected as a modular component rather than a monolithic product \cite{caddle2025building}. This involves designing open APIs or plugins that allow the risk detection engine to be integrated into established school or non-profit organizations without ceding governance. Such ``dual-use'' architectures should include hard-coded ethical guardrails—such as open-source auditing logs—to ensure that integration does not compromise the tool's teen-centric values for the sake of institutional monitoring \cite{Ekambaranathan2023HowChallenges}.

\expert{\subsection{Implications for Experts as Secondary Stakeholders: Aligning Interests for Teen Online Safety}
\textbf{Realigning online safety metrics with teen or family needs requires a shift from ``policing'' to ``advocacy.''} Our findings reveal a salient misalignment between the technical capabilities of risk detection technology and the perceived supportive needs of teens. While industry professionals in our study prioritized the accuracy of detection (Section \ref{tension2}), practitioners emphasized that detection without meaningful intervention is merely ``policing'' teens rather than supporting them. This echoes prior HCI research on the failure of restrictive parental controls, which often erode trust without improving safety outcomes \cite{Wisniewski2017ParentalSafety, Ghosh2018SafetyControl}. To move from a surveillance model to an advocacy one, secondary stakeholders, specifically developers and policymakers, must shift their success metrics of online safety from \textit{detection rates} to \textit{resolution rates}. That is, instead of simply flagging a ``risk,'' platforms should be evaluated on their ability to bridge the ``intervention gap'' identified by our participants (Section \ref{tension2}), offering evidence-based pathways to trusted adult allies. This shifts the role of risk detection technology from a monitor that exposes teens to a coach that empowers them, mitigating the privacy distrust (Section \ref{tension4}) that experts warned would lead to teens' avoidance.

\textbf{Regulatory debates must resolve the paradox between equitable access and financial sustainability.} Our experts identified a conflict between the trustworthiness of non-profit models for online safety solutions and the financial reality of sustaining them (Section \ref{tension5}), warning that pay-to-play safety solutions would privilege affluent youth while leaving marginalized teens exposed. This finding extends prior work on the digital divide in youth online safety \cite{Razi_Sexual, badillo2024towards}, suggesting that current market-driven safety solutions may inadvertently deepen these inequities. Thus, regulation that mandates safety solutions without providing the financial infrastructure to sustain them inadvertently promotes a for-profit compliance industry that prioritizes liability protection over teen well-being. To resolve this, secondary stakeholders should mandate commercial platforms to fund independent, third-party safety layers, like our MOSafely, without exerting control over the data or governance. This decouples the \textit{cost} of safety paid by social media platforms from the \textit{incentive} of safety managed by trusted non-profits, ensuring that ``mature'' safety tools are accessible to all demographics, not just those who can afford a subscription.

\textbf{Integrating safety solutions into communities is necessary to operationalize a ``mature safely'' paradigm.} Finally, our participants argued that a standalone technical tool, no matter how well-designed, will fail if it relies solely on a teen's intrinsic motivation, which is often affected by social stigma (Section \ref{tension3}). This aligns with Boyd's characterization of networked publics \cite{boyd2014s}, where social acceptance can supersede privacy or safety concern. Practical improvement in teen safety requires secondary stakeholders, such as practitioners and clinicians, to move beyond passive endorsement and actively integrate these tools into existing communities \cite{Caddle2024AOnline}. Rather than treating risk detection as a private, individual task, schools and youth organizations could frame the use of such dashboards as a component of digital literacy curricula, normalizing their use to reduce the stigma factor. By embedding the tool within a broader community context, as suggested by our participants who called for ``community buy-in'' (Section \ref{tension3}), secondary stakeholders can create the social scaffolding necessary for teens to engage with the risk detection technology.}

\section{LIMITATION \& FUTURE WORK} 
Our study has limitations informing future work. First, we focused on expert \methods{participants who are secondary stakeholders to youth online safety}, rather than directly involving teens in our study. This was a thoughtful choice to first ``mature'' the technological concept and \methods{use expert feedback to identify and mitigate potential risks} before engaging with vulnerable teens. Future work will be cognitive walkthrough and interview sessions with teens, using our findings to \novelty{validate whether these expert-identified tensions manifest in practice. For instance, guided by expert concerns regarding privacy distrust (Section \ref{tension4}), our future protocol will implement a staged consent to build trust before data collection. Similarly, to address the need for meaningful intervention (Section \ref{tension2}), we will shift our evaluation metrics from simple detection accuracy to the perceived utility of supports.} Second, our experts reacted to a video demo with simulated data, not a live system. \methods{As detailed in Section 3.2, we chose this approach to prioritize privacy and responsibly gather feedback on the MOSafely design concept before exposing teens and their private data to it.}

\section{CONCLUSION}
In this study, we presented an expert evaluation of our teen online safety solution, MOSafely, a teen-centered risk detection dashboard, to uncover the challenges of implementing and sustaining such technologies. Our analysis of interviews with 33 online safety experts identified five primary tensions, including the foundational conflict between objective risk definitions and teens’ context-dependent realities, and the disconnect between the goal of teen empowerment and their intrinsic motivation. Our findings call for rethinking the ``teen-centered'' approach to be involve and align more adult stakeholder for youth online safety, addressing the intertwined technical, social, and political-economic misalignments between a risk detection technology’s design and its real-world context. Our findings also demonstrate how expert evaluation can help enable a shift from a ``fail fast'' to a ``mature \clarity{safely}'' paradigm for risk detection technologies, \clarity{allowing researchers or developers to responsibly identify and address practical concerns before deployment, rather than discovering them after the technologies cause harm to youth. Ultimately, to truly protect teens, we must design not just for them but for the real-world context they inhabit, ensuring our solutions are as effective as they are sustainable.}

\begin{acks}
We thank the Associate Chairs and anonymous reviewers for their constructive feedback. We also gratefully acknowledge Xavier Caddle for his efforts in data collection and his contributions to this research, and we thank all participants for their time and contributions. Dr. Wisniewski's research is supported by the U.S. National Science Foundation under grants \#TI-2550746, \#CNS-2550834, and \#IIS-2550812, and by the William T. Grant Foundation grant \#187941. Any opinions, findings, and conclusions or recommendations expressed in this material are those of the authors and do not necessarily reflect the views of the research sponsors.
\end{acks}

\bibliographystyle{ACM-Reference-Format}
\bibliography{main}

\end{document}